# Finite time analysis based on Sum of Squares Technique: Applied to the super-twisting second order sliding mode control [○]

Sina Sanjari[a], Sadjaad Ozgoli[*b]

[a] Tarbiat modares university and s.sanjari@modares.ac.ir
[*] corresponding author. Phone +98 21 8288 3309  Fax +98 21 8288 4325
[b] Tarbiat modares university and ozgoli@modares.ac.ir

**Abstract**

Finite time analysis of the continuous system is investigated through both stability and stabilization based on Sum of squares programming. A systematic approach is proposed to construct Lyapunov function and Control Lyapunov function for this objective. The Region of reaching, the set which has the property that all trajectories starting from initial point inside it reach to the origin in finite time, is introduced, and The largest subset of region of reaching is estimated using Lyapunov based technique. The main results are presented to give sufficient conditions which can be translated by semi-definite program. These conditions are provided a feasibility problem involving sum of squares constraints. The results of the paper are then verified by several simulation and numerical examples. Furthermore, one important practical application namely as a super twisting second order sliding mode control also is presented using the proposed result to illustrate its effectiveness.

## 1. Introduction

Finite time stability, i.e. the stability in the sense of reaching to the origin in finite time, has been the center of attention in many recent articles. It was shown that systems include finite time stability property have faster convergence rate and also better robustness against disturbances. Much work has been done in the literature to enrich the theoretical aspect of this concept; the finite time stability and stabilization problem are addressed for second-order systems in [1, 2]; these results have been further extended for a class of controllable systems [3] and a class of systems with a special triangular form [4]; an approach based on Lyapunov function technique has been introduced to facilitate convergence analysis and to construct finite-time controller for a class of continuous systems using control Lyapunov function (CLF) [5]; a general framework has been developed for finite time stabilization using vector Lyapunov functions  [6, 7]; the method proposed in [8], makes use of Holder continuous state feedback to guarantee global finite time stability for a family of uncertain nonlinear systems, and finally, finite time stabilization by means of output feedback has been studied in [9].

From a practical point of view, finite time stability (FTS) has advantages over the classical Lyapunov stability. FTS controller has proven to be applicable to robot systems through either state feedback or dynamic output feedback [10, 11], to stochastic systems [12, 13] and to impulsive dynamical systems [14]. Nevertheless, despite the rapid advancements in theory, there is still a huge gap between theory and practice of finite time stability/stabilization. In this regard, introducing a systematic approach seems necessary to alleviate the problem.

The objective of this paper is to investigate finite-time stability and stabilization problems for a class of continuous systems that satisfy uniqueness of solution only in forward time. These tasks are performed by employing an algorithmic approach based on Lyapunov functions. It will be demonstrated that the absence of Lipschitzian property, a conclusion from non-uniqueness of solutions in backward time, is essential for the aforesaid systems to be finite time stable. An important merit of this approach is in that a Lyapunov function ensuring finite time stability is found based on a systematic procedure. Moreover, sufficient conditions that provide finite time stability and stabilization are transformed into sum of squares equalities and inequalities which can be checked systematically. Next, CLF-based finite time controller is proposed in order to achieve finite time stability of the closed loop dynamics. As an example, the proposed approach is applied to one of the most important types of Second order sliding mode control [15] known as "super twisting" to demonstrate its effectiveness.

Note that the SOS technique is the foundation of the algorithms developed in this paper. This technique is origi-nally used for systems with polynomial or rational vector fields, but thanks to its extension to non-polynomial systems [16, 17], its applicability has also been significantly extended. In fact, non-Lipschitzian property posits that system dynamics cannot be a polynomial and contain non-polynomial terms (regularly rational power terms) at the same time; accordingly, one of the methods introduced in [16, 17] is utilized to enable SOS technique to treat non-polynomial systems.

The rest of the paper is organized as follows: Section 2 presents preliminaries on SOS optimization. Section 3 discusses the mentioned control problem, and presents the general dynamics of target systems. SOS-based programming, used for finite time analysis, is then presented in Section 4. In Section 5, region of reaching is introduced and a lemma is proposed to obtain this set. Then the largest region set has been estimated using SOS technique. Finite time stabilization problem is investigated in Section 6. In Section 7, three numerical examples are given to show the effectiveness of the presented method. Applications of the proposed method in the super twisting algorithm are given in Section 8. Finally, Section 9 concludes the paper.

## 2. Mathematical preliminaries

This section presents a brief review on SOS decomposition and other definitions needed to follow the paper.

**Definition 1 (monomial).** Let $\mathcal{R}[x] = \mathcal{R}[x_1 \ldots, x_n]$ be the set of real polynomials with real coefficients. A function

$$Z_\alpha = x_1^{\alpha_{i_1}} x_2^{\alpha_{i_2}} \ldots x_n^{\alpha_{i_n}}$$

With $\{\alpha_{i_1}, \ldots, \alpha_{in}\} \in Z_+$ is called a monomial and its degree is given by $\deg(Z_\alpha) = \sum_{i=1}^n \alpha_i$.

**Definition 2 (polynomial).** A real polynomial function $p \in \mathcal{R}[x]$ is defined as

$$p(x) = \sum_k C_k Z_{\alpha_k}$$

where $C_k \in \mathbb{R}$ and $x \in \mathbb{R}^n$. The polynomial $p(x)$ is said to be of degree $m$ if it corresponds to the largest monomial degree in $p(x)$ i.e. $m = max_k \deg(Z_{\alpha k})$.

**Definition 3 (SOS).** A real polynomial $p(x) \in \mathcal{R}_n$ which has a degree of $d$ is SOS if there exist polynomials such that

$$p(x) = \sum_{i=1}^r p_i^2(x)$$

The subset of all SOS polynomials in $\mathcal{R}_n$ is denoted by $\Sigma_n$.

The SOS definition implies that the existence of SOS decomposition is sufficient condition for $p(x)$ to be positive semi-definite, i.e. $p(x) \geq 0$. In general the converse of this result does not hold; however, the possibility of $\mathcal{R}_n$ being $\Sigma_n$ has been calculated in [16]. It is demonstrated that the gap between these two sets is negligible.

In most control problems, "Lyapunov problem" for example, it is important to investigate the non-negativity of polynomials. In general, it is extremely hard or sometimes even impossible to solve this problem. However, to check whether a polynomial is sum of squares or not, is a Semi-definite programming (SDP) and can be easily done. So in our problem formulation, conditions on non-negativity are replaced by sufficient conditions for the polynomial to be SOS.

**Lemma 2 (generalized S-procedure [16]):** Given set of polynomials $\{p_i\}_{i=0}^m \in \mathcal{R}_n$, if there exists $\{s_k\}_{i=1}^m \in \Sigma_n$ such that $p_0 - \sum_{i=1}^m s_i p_i \in \Sigma_n$, then

$$\cup_{i=1}^m \{x \in \mathbb{R}^n | p_i(x) \geq 0\} \subseteq \{x \in \mathbb{R}^n | p_i(x) \geq 0\}$$

This lemma is a generalization of the well-known S-procedure.

**Notation:** For a polynomial $Q \in \mathcal{R}$, $Q \geq 0$ (respectively $Q > 0$) represents positive semi-definiteness (respectively positive definite) of $Q$; $\mathcal{R}_n[x]$ denotes the set of all polynomials in $n$ variables; $\Sigma_n$ denotes the set of all sum of squares polynomials in n variables; $CL^0(\mathcal{E}, \mathcal{F})$ (respectively $CL^k(\mathcal{E}, \mathcal{F})$) denotes the set of all continuous functions on $\mathcal{E}$, locally Lipschitz on $\mathcal{E} \setminus \{0\}$ with value in $\mathcal{F}$ (respectively the set of continuous functions on $\mathcal{E}$, $C^k$ on $\mathcal{E} \setminus \{0\}$ with value in $\mathcal{F}$); $\mathcal{B}^n$ is the unit ball in $\mathbb{R}^n$.

## 3. System description and problem statement

Consider the following nonlinear continuous system as:

$$\dot{x} = f(x) \tag{1}$$

In which $x \in \mathbb{R}^n$, and $f \in C^0(\mathbb{R}^n)$. Without loss of generality, assume the origin is an equilibrium point of this system. Equivalently

$$f(0) = 0 \tag{2}$$

**Assumption1.** The function $f$ is included in the class $CL^k(\mathbb{R}^n, \mathbb{R}^n)$-functions for all $k \geq 0$ .i.e. the set of continuous functions on $\mathbb{R}^n$, and $C^k$ on $\mathbb{R}^n \setminus \{0\}$ with value in $\mathbb{R}^n$.

**Remark1.** Since the class of systems with uniqueness of solutions in forward time is included in the class $CL^k$-systems, assumption 1 provides sufficient condition for existence of such kind of solutions.
In the following the definition of finite time stability introduced in [5] is given.

**Definition 4.** The origin is finite time stable for the system (1) if there exists a nonempty neighborhood $\mathcal{V} \in \mathbb{R}^n$ of the origin such that:
(1) There exists a function $T: \mathcal{V} \setminus \{0\} \to \mathbb{R}_{\geq 0}$ such that if $x_0 \in V \setminus \{0\}$ then $\varphi^{x_0}(t)$ is defined (and particularly unique) on $[0, T(x_0)[, \varphi^{x_0}(t) \in V \setminus \{0\}$ for all $t \in [0, T(x_0)[$ and $\lim_{t \to T(x_0)} \varphi^{x_0}(t) = 0$ is called the settling-time of the system (1);
(2) For all $\varepsilon > 0$, there exists $\delta(\varepsilon) > 0$, for every $x_0 \in (\delta(\varepsilon) \mathcal{B}^n \setminus \{0\}) \cap V, \varphi^{x_0}(t) \in \varepsilon \mathcal{B}^n$ for all $t \in [0, T(x_0)[$.
By this definition we can conclude that the solution $\varphi^x$ reach to the origin in finite time, and since system is autonomous, it ensures that the solution of system remain at the origin for the rest of time. Therefore, settling time is obtained as:

$$T(x) = \inf\{t \in \mathbb{R}_{\geq 0} : \varphi(t, x) = 0\} \quad (3)$$

The primary objective of the reminder of this work is to propose a method that obtains Lyapunov function systematically based on SOS framework ensuring finite time stability both locally and globally for this class of system. Next, a control law based on Lyapunov method to guarantee finite time stability accomplished accordingly.

## 4. Finite time stability analysis

This section concentrated on finding Lyapunov function which gives sufficient condition for finite time stability. For this purpose, two algorithmically approaches based on SOS method proposed.

The following lemma gives necessary and sufficient condition for finite time stability.

**Lemma 3** [5]. Consider system (1) satisfying assumption 1, the origin is finite time stable with class $CL^0$ settling function if and only if there exists a real number $c > 0$, $\alpha \in [0,1)$ and a class $CL^\infty$-positive definite Lyapunov function $V$ satisfying the following condition :

$$\dot{V}(x) \leq -cV^\alpha(x) \quad (4)$$

Moreover, settling function $T(x)$ satisfies the below inequality.

$$T(x) \leq \frac{V(x)^{1-\alpha}}{c(1-\alpha)} \quad (5)$$

Motivated by lemma 1 we propose two algorithmically method for derivation of sufficient condition for finite time stability using sum of squares.

**Theorem 1.** System (1) satisfying assumption (1) is finite time stable if the unknown Lyapunov function is found which is constructed by

1) *Choosing small constants $\varepsilon_{ij}$ and constructing*

$$l_k(x) = \sum_{i=1}^{n}\sum_{j=1}^{d} \varepsilon_{ij} x_i^{2j}, \sum_{j=1}^{m}\varepsilon_{ij} > 0, \forall i,j = 1,\dots,n, \varepsilon_{ij} \geq 0, k = 1 \quad (6)$$

2) *Solving the following SOS program*

---
Find $V \in \mathcal{R}_n, V(0) = 0$ for a given positive scaler c

Such that

$$V - l_1 \in \Sigma_n \quad (7)$$

$$-(\frac{\partial V}{\partial x}f(x) + c) \in \Sigma_n \quad (8)$$
---

Moreover, the settling time satisfies the condition:

$$T(x) \leq \frac{V(x)}{c} \quad (9)$$

**Proof.** According to lemma 3, the problem can be posed as an emptiness form of set containment constraint lemma 3

$$\{x \in \mathbb{R}^n \mid V(x) \leq 0, x \neq 0\} = \emptyset \quad (10)$$

$$\{x \in \mathbb{R}^n \mid (\frac{\partial V}{\partial x}f(x) + cV^\alpha(x)) > 0, (\frac{\partial V}{\partial x}f(x) + cV^\alpha(x)) \neq 0\} = \emptyset \quad (11)$$

By introducing fixed positive definite polynomials $l_i(x)$, replacing $x \neq 0$ with $l_i(x) \neq 0$ then Applying P-satz lemma, the constraint (10) and (11) are rewritten as:

$$s_1 - s_2 V + l_1^{2k_1} = 0 \quad (12)$$

$$s_3\left(\frac{\partial V}{\partial x}f(x) + cV^\alpha(x)\right) + s_4 + (\frac{\partial V}{\partial x}f(x) + cV^\alpha(x))^{2k_2} = 0 \quad (13)$$

In the above $s_1, \dots, s_4 \in \Sigma_n$ and $k_1, k_2 \in \mathbb{Z}$. Generally, these constraints cannot be checked with SOS programming. In order to make the problem solvable by SOS programming we choose $\alpha = 0$, and $k_1 = k_2 = 1$. Additionally, since the product of two SOS polynomials is also SOS, we can restate the constraint as in (7) and (8); so, the proof is completed. ∎

**Remark 2.** Regarding to the definition of finite time stability, if the origin of system is finite time stable, $f(x)$ cannot be locally Lipschitz at the origin, and contains non-polynomial terms (rational power terms). Therefore, to solve this problem, we use the recasting procedure (see [16]) to transform non-polynomial system by defining slack variables. Thus, constraints (7) and (8) are restated as follows:

$$V - l_1(\tilde{x}_1, \tilde{x}_2) - \lambda_1^T(\tilde{x}_1, \tilde{x}_2)G_1(\tilde{x}_1, \tilde{x}_2) - \sigma_1^T(\tilde{x}_1, \tilde{x}_2)G_2(\tilde{x}_1, \tilde{x}_2) \in \Sigma_n \quad (14)$$

$$-(\frac{\partial V}{\partial x}f(x) + c - \lambda_2^T(\tilde{x}_1, \tilde{x}_2)G_1(\tilde{x}_1, \tilde{x}_2) - \sigma_2^T(\tilde{x}_1, \tilde{x}_2)G_2(\tilde{x}_1, \tilde{x}_2)) \in \Sigma_n \quad (15)$$

$\tilde{x}_1$ contains the original variables of system, and $\tilde{x}_2$ includes the slack variables. Polynomial column vectors $\lambda_1(\tilde{x}_1, \tilde{x}_2), \lambda_2(\tilde{x}_1, \tilde{x}_2)$ and sum of squares polynomial vectors $\sigma_1(\tilde{x}_1, \tilde{x}_2), \sigma_2(\tilde{x}_1, \tilde{x}_2)$ are considered of appropriate dimensions that satisfy

$$G_1(\tilde{x}_1, \tilde{x}_2) = 0 \quad (16)$$

$$G_2(\tilde{x}_1, \tilde{x}_2) \geq 0 \quad (17)$$

**Remark 3.** According to assumption $\alpha = 0$, the previous theorem poses a conservative condition which is indeed tight between all $\alpha \in [0,1)$. The following theorem is therefore gives a solution with gives milder condition compared to theorem 1 so that conservativeness of (8) will be relaxed.

**Theorem 2.** System (1) satisfying assumption 1 is finite time stable if the unknown Lyapunov function is found which is constructed by

1) Choosing small constants $\varepsilon_{ij}$ and constructing $l_k(x)$ as (6)
2) Solving the following SOS program

Find $V \in \mathcal{R}_n, V(0) = 0$ and $s_2, s_4, s_8$ for a given positive scaler c

$$V - l_1 \in \Sigma_n \tag{18}$$

$$-((1-V)s_4 + s_2(\frac{\partial V}{\partial x}f(x) + cV) + s_8 + l_2) \in \Sigma_n \tag{19}$$

$$-((V-1)s_9 + s_{10}(\frac{\partial V}{\partial x}f(x) + c)) \in \Sigma_n \tag{20}$$

**Proof.** Define the set $\Omega_1 \triangleq \{x \in \mathbb{R}^n | V \leq 1\}$, and pose the problem as

$$V(x) > 0, \forall x \in D - \{0\} \tag{21}$$

$$\{x \in \mathbb{R}^n | V(x) \leq 1\} \setminus \{0\} \subseteq \{x \in \mathbb{R}^n | \frac{\partial V}{\partial x}f(x) \leq -cV^\alpha\} \tag{22}$$

According to the assumption set $\Omega_2$, for $\alpha \in [0,1)$

$$V^\alpha - V = b(x)$$

where $b(x)$ is a positive definite function, and is enhanced when $V(x)$ is decreased. In other words, it changes inversely with respect to changes in $V(x)$. The constraint (22) can be rewritten as

$$\{x \in \mathbb{R}^n | V(x) \leq 1\} \setminus \{0\} \subseteq \{x \in \mathbb{R}^n | \frac{\partial V}{\partial x}f(x) \leq -cV - b(x)\} \tag{23}$$

Now according to the properties of $b(x)$ must possess, one good choice can be $b(x) = 1/l_3$. Thus, the problem can be posed as an emptiness form of set containment constraint

$$\{x \in \mathbb{R}^n | V(x) \leq 0, x \neq 0\} = \emptyset$$

$$\{x \in \mathbb{R}^n | V(x) \leq 1, \frac{\partial V}{\partial x}f(x) + cV + 1/l(x) \geq 0, \frac{\partial V}{\partial x}f(x) + cV + 1/l(x) \neq 0, x \neq 0\} = \emptyset$$

Applying P-satz lemma, constraints can be rewritten as follows:

$$s_1 - s_2 V + l_1^{2k_1} = 0$$

$$s_3 + s_4(1-V) + s_5\left(\frac{\partial V}{\partial x}f(x) + cV + 1/s_6\right)$$
$$+ s_7(1-V)\left(\frac{\partial V}{\partial x}f(x) + cV + \frac{1}{s_6}\right)$$
$$+ \left(\frac{\partial V}{\partial x}f(x) + cV + \frac{1}{s_6}\right)^{2k_2} + l_2^{2k_3} = 0$$

Similar to the proof of theorem 1, we can restate constraints as (18) and (19) by simplification; using same cal-culation in theorem 1 for the set $\Omega_2 \triangleq \{x \in \mathbb{R}^n | V > 1\}$ constraint (20) is concluded; therefore, the proof is completed. ∎

**Remark 4.** Since constraint (18) has a term containing a product of decision variables, this constraint cannot be checked by a linear semi-definite programming, but can be converted to a bilinear semi-definite program which is solvable by aid of PENBMI solver [18], a local bilinear matrix inequality (BMI) solver from PENOPT, or by using iterative algorithm.

Although theorem 2 assuage conservativeness problem of programming but it due to complexity of computation approach. The following corollary relaxed this complexity when a Lyapunov fuction is restricted to quadratic function.

**Corollary 1.** Consider system (1) satisfying assumption 1. If the following SOS program is feasible then, the origin is finite time stable for system is introduced in (1).

max $c$ over matrix $Q$, for a given vector $w(x)$ and a positive constant $\varepsilon_1$ and positive integers $q > p$

such that

$$(Q - \varepsilon_1 I) \in \Sigma_n \tag{24}$$

$$-(\frac{\partial w}{\partial x}f(x)Qw(x) + w(x)^T Qf(x)\frac{\partial w}{\partial x} + cM) \in \Sigma_n \tag{25}$$

With constraint

$$M^q = (w(x)^T w(x))^p \tag{26}$$

**Proof.** (24) concludes that matrix $Q$ is positive definite. Defining the Lyapunov function using generalized krasovskii idea, (24) implies that Lyapunov candidate function is positive definite i.e.

$$V = w(x)^T Q w(x) > 0$$

Where $w(x)$ is chosen such that . Taking derivative of the Lyapunov function with respect to time and using (24) and (25) concludes that

$$\dot{V} = \frac{\partial w}{\partial x}f(x)Qw(x) + w(x)^T Qf(x)\frac{\partial w}{\partial x} \leq c\|w(x)\|^{2p/q} \tag{27}$$

since

$$\|w(x)\|^2 = \left(\sqrt{Q}w(x)\right)^T Q^{-1}\left(\sqrt{Q}w(x)\right) \geq$$
$$\lambda_{min}(Q^{-1})\left\|\left(\sqrt{Q}w(x)\right)\right\|^2 = \lambda_{min}(Q^{-1})V \tag{28}$$

It can be easily shown that (27) and (28) yields that constraint (4) is satisfied for $\alpha = (p/q) < 1$. ∎

**Remark 5.** Although selecting a right $w(x)$ maybe in some case be difficult, but it is useful in the way that the designer can consider some inside structural information about Lyapunov function into a design approach; therefore, analytic reasoning and efficient computing can be consid-ered in this design method. For example one choice for $w(x)$ is:

$$w(x) = \begin{bmatrix} 1 & 0 & \cdots & 0 \\ L_{11}(x) & 1 & 0 & 0 \\ \vdots & & \ddots & 0 \\ L_{n1}(x) & L_{n2}(x) & \cdots & 1 \end{bmatrix} x$$

However, in order to simplify a computing $w(x)$ can be considered as only $x$, or $f(x)$.

## 5. Estimation of Region of Reaching

In this section, the largest set with the property that all trajectories starting from initial point inside it reach to the origin in finite time is estimated using SOS optimization problem. For this purpose, we use the following lemma which is a modification of ideas from [19] and [20].

**Lemma 4.** If there exists a Lyapunov function $V : \mathcal{V} \to R$ such that

$$V > 0 \tag{29}$$

$$\Omega := \{x \in \mathbb{R}^n \mid V(x) \leq 1\} \text{ is bounded set} \tag{30}$$

$$\{x \in \mathbb{R}^n \mid V(x) \leq 1\} \setminus \{0\} \subseteq \{x \in \mathbb{R}^n \mid \frac{\partial V}{\partial x} f(x) \leq -cV^\alpha\} \tag{31}$$

Then for all $x(t_0) \in \Omega$, the solution of (4) reach to the origin in finite time. As such, $\Omega$ is a subset of the region of attraction for (4).

**Proof.** Condition (29) concludes that if $x_0 \in \Omega$, then $V(x) \leq V(x_0) \leq 1$. This means that solutions $\varphi^{x_0}$ in $\Omega$ exists and remains inside $\Omega$. Define the set $S_\varepsilon = \{x \in \mathbb{R}^n \mid \frac{\varepsilon}{2} \leq V(x) \leq 1\}$ for a taken $\varepsilon > 0$. According to (24) $S_\varepsilon \subseteq \Omega \setminus \{0\} \subseteq \{x \in \mathbb{R}^n \mid \frac{\partial V}{\partial x} f(x) \leq -cV^\alpha\}$ for $\alpha \in [0,1)$. Since $S_\varepsilon$ is a compact set, $\exists r_\varepsilon > 0$ such that $\dot{V}(x) - cV^\alpha \leq r_\varepsilon < 0$ on $S_\varepsilon$; therefore, $\exists t^*$ such that $V(x) < \varepsilon$ for all $t > t^*$. This implies that if , then $V(x) \to 0$ as $t \to \infty$.

Let $\varepsilon > 0$, Define $\Omega_\varepsilon = \{x \in \mathbb{R}^n \mid \|x\| \geq \varepsilon, V(x) \leq 1\}$. $\Omega_\varepsilon$ is compact, with $0 \notin \Omega$. Since $V$ is continuous and positive definite according to (29), $\exists \gamma$ such that $V(x) \geq \gamma > 0$ on $\Omega_\varepsilon$. We have already established that $V(x) \to 0$ as $t \to \infty$. So, $\exists \hat{t}$ such that for all $t > \hat{t}$, $V(x) < \gamma$, and hence , $x \notin \Omega_\varepsilon$, which implies that the origin of system is asymptotically stable for $x_0 \in \Omega$. Define a function $[0,T(x)) \to (0,V(x)]$, $t \to V(\varphi(x,t))$. Since this function is decreasing and differentiable, its inverse $(0,V(x)] \to [0,T(x))$, $s \to \theta_x(s)$ is differentiable and satisfies the following condition for all $s \in (0,V(x)]$.

$$\theta'_x(s) = \frac{ds}{-\dot{V}(\varphi(\theta_x(s),x))}$$

Using the variable change $s = V(\varphi(x,t))$ concludes that

$$T(x) = \int_0^{T(x)} dt = \int_{V(x)}^0 \theta'_x(s) ds = \int_0^{V(x)} \frac{ds}{-\dot{V}(\varphi(\theta_x(s),x))} < +\infty$$

Therefore, this implies that the origin of system is finite time stable for $x_0 \in \Omega$. ∎

**Theorem 3.** Consider system (1) satisfying assumption 1. All solution starting in $\Omega$, which is a subset of the region of reaching for (1), reach to the origin in finite time if the following SOS optimization program is feasible.

$$\max \beta \text{ over } V \in \mathcal{R}_n, V(0) = 0, s_6, s_8 \in \Sigma_n,$$

such that

$$V - l_1 \in \Sigma_n \tag{32}$$

$$-((\beta - p)s_6 + (V - 1)) \in \Sigma_n \tag{33}$$

$$-((1 - V)s_8 + \frac{\partial V}{\partial x} f(x) + cV + l_2) \in \Sigma_n \tag{34}$$

**Proof.** Define fixed variable-sized region $p_\beta = \{x \in \mathbb{R}^n \mid p(x) \leq \beta\}$ whose shape is determined by $\beta$. Considering $p(x)$ a positive definite polynomial and maximizing $\beta > 0$ enlarge the set $\Omega$. Thus, the problem of finding this region can be posed as the following optimization problem.

$$\max_{V \in \mathcal{R}_n} \beta \ s.t. V(x) > 0 \text{ for all } x \in \mathbb{R}^n \setminus \{0\} \text{ and } V(0) = 0 \tag{35}$$

$$\text{the set } \{x \in \mathbb{R}^n \mid V(x) < 1\} \text{ is bounded} \tag{36}$$

$$\{x \in \mathbb{R}^n \mid p(x) \leq \beta\} \subseteq \{x \in \mathbb{R}^n \mid V(x) < 1\} \tag{37}$$

$$\{x \in \mathbb{R}^n \mid V(x) < 1\} \setminus \{0\} \subseteq \{x \in \mathbb{R}^n \mid \frac{\partial V}{\partial x} f(x) + cV < 0\} \tag{38}$$

Expressing set of containment above constraint , and Applying P-satz lemma similar to the proof of theorem 1 and theorem 2 complete the proof. ∎

**Remark 6.** Similar to remark 3, we can utilize PENBMI solver or iterative algorithms in order to solving BMI constraints (33), (34); however, both this methods have own drawback Such as local solvability instead globally for PENBMI, and dependence to initial search for iterative method. On the other hand it can be proved similar to the one presented in [21] that this type of problem can converted to a quasi-convex problem, and can be solved by this approach.

## 6. Finite time stabilization

This section concentrates on global and local finite time stabilization for class of affine continuous systems by means of SOS programming. For this purpose Modified Sontag feedback law [22] finite time stabilizing system based on control Lyapunov function [5] is utilized and be solved given helped by SOS approach.
Consider the affine system:

$$\dot{x} = f_0(x) + \sum_{i=1}^m f_i(x) u_i \tag{39}$$

Where $x \in \mathbb{R}^n, u \in \mathbb{R}^m$ and $f_i \in CL^k(\mathbb{R}^n, \mathbb{R}^n)$ for all $0 \leq i \leq m$ and $f_0(0) = 0$.

The objective is to designing control signal $u$ such that system represented by (39) becomes finite time stable. Accordingly, the following theorem proposes an SOS-based approach to synthesize a control law that ensures finite time stability in the global sense.

**Theorem 4.** Introduced system (39) will be globally finite time stable by applying the following control law.

$$u(x) = \begin{cases} -b_i(x) \frac{a(x)+\sqrt[p]{a(x)^p+\beta(x)^q}}{\beta(x)} & \beta(x) = 0 \\ 0 & \beta(x) \neq 0 \end{cases} \quad (40)$$

Where $p$ and $q$ are even integers and $p, q \geq 2$. $b_i(x) = \frac{\partial V}{\partial x} f_i(x)$, $a(x) = \left(\frac{\partial V}{\partial x} f_0(x)\right)$, $\beta(x) = \sum_{i=1}^{m} \left(\frac{\partial V}{\partial x} f_i(x)\right)^2$.

And The unknown function $V$ is constructed by

3) Choosing small constants $\varepsilon_{ij}$ and constructing $l_k(x)$ as (6)

4) Solving the following SOS program

Find polynomials $V(x)$, $p_{2i}$ and SOS polynomial $s_1$ and, positive scalar c such that

$$V - l_1 \in \Sigma_n \quad (41)$$

$$s_1\left[\frac{\partial V}{\partial x} f_0(x) + c\right] + \sum_{i=1}^{m} p_{2i}\left[\frac{\partial V}{\partial x} f_i(x)\right] + l_2 \in \Sigma_n \quad (42)$$

**Proof.** The proof can be concluded using the same technique in theorem 1 and [5]. ∎

In order to extend the scheme given in theorem 4 to local finite time stability, the next theorem synthesizes a corresponding control law, and calculates the parameters straightforward by solving an SOS optimization problem.

**Remark 7.** The same strategy as the one used in theorem 2 can be hired for conservativeness issue in theorem 4.

**Theorem 5.** The system in (39) will be locally finite time stable by applying (40) in which the unknown function is constructed by $V(x)$ is constructed by
1) Choosing small constants $\varepsilon_{ij}$ and constructing $l_k(x)$ as (6)
2) Solving the following SOS optimization problem

Max $\beta$ over $s_1, s_2, s_8 \in \Sigma_n, V, p_4 \in \mathcal{R}_n, V(0) = 0$

such that

$$V - l_1 \in \Sigma_n \quad (43)$$

$$-((\beta - p)s_8 + (V - 1)) \in \Sigma_n \quad (44)$$

$$-\{s_1(1 - V) + s_2[\frac{\partial V}{\partial x} f(x) + cV] + p_4 \frac{\partial V}{\partial x} g(x) + l_2\} \in \Sigma_n \quad (45)$$

**Proof**: the proof can be concluded using the same technique in theorem 2 and [5]. ∎

### 7. Illustrative examples

In this section, some examples are provided to show the applicability and flexibility of the method developed in this paper. It should be noted that anywhere needed, the SOS programs are solved by means of the SOSTOOLS.

**Example 1.** Consider the following system which satisfies assumption (1) as:

$$\begin{cases} \dot{x}_1 = -x_1^{\frac{1}{3}} - x_1^3 + x_2 x_1^{\frac{1}{3}} \\ \dot{x}_2 = -x_2^{\frac{1}{3}} - x_2^3 - x_1 \end{cases}$$

Considering an arbitrary positive scalar $c$ (take $c = 1$ for example), theorem 1 gives a Lyapunov function guaranteeing finite time stability of the aforementioned system:

$$\begin{aligned} V = {} & 0.6615x_1^4 - 0.01586x_1^3 x_2 - 0.1176x_1^3 \\ & + 0.9933x_1^2 x_2^2 + 0.6459x_1^2 x_2 \\ & + 1.389x_1^2 + 0.1926x_1 x_2^3 \\ & - 0.08476x_1 x_2^2 + 0.08316x_1 x_2 \\ & + 0.6821x_2^4 - 0.03822x_2^3 + 1.326x_2^2 \end{aligned}$$

Finite time stability of the system is manifested in Fig (1).

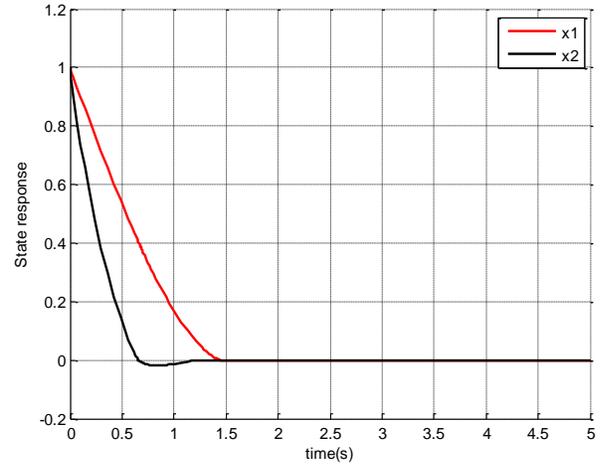

Fig. 1. State trajectories example 1

The following Lyapunov function is obtained by exploiting theorem 2.

$$V = 4.945x_1^2 + 1.159x_1 x_2 + 4.494x_2^2$$

As seen, the Lyapunov function obtained from theorem 2 has a lower degree in comparison with the one achieved by using theorem 1. This shows that theorem 2 offers less conservative solutions.

**Example 2.** This example illustrates the effectiveness of the approach in obtaining finite time controller. Consider the system

$$\begin{cases} \dot{x}_1 = -|x_1|^{\frac{1}{8}}sign(x_1) - x_2 \\ \dot{x}_2 = |x_1|^{\frac{1}{8}}sign(x_1)|x_2|^{\frac{7}{8}} + |x_2|^{\frac{1}{8}}u \end{cases}$$

, and define the below slack variables to transform this system into a polynomial one:

$$\begin{cases} x_3 = |x_1|^{\frac{1}{8}} \\ x_4 = |x_2|^{\frac{1}{8}} \end{cases}$$

Applying theorem 4, and taking remark 1 into account, we get the Control Lyapunov Function as:

$$V = 0.9861x_3^9 + 0.9744x_4^9$$

Eventually, the corresponding control law which stabilizes the system (39) in finite time sense is

$$u = \frac{x_3^2 - \sqrt[4]{x_3^8 + x_4^8}}{\frac{x_2}{x_4^6}}$$

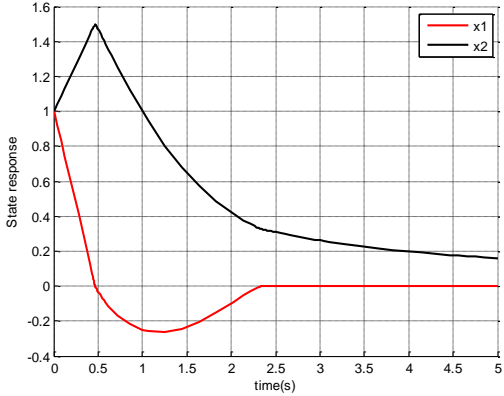

Fig. 2. Open loop trajectories example 2

Fig.2 depicts the open-loop system trajectories, and Fig.3 shows the closed-loop system trajectories when the control law (40) is applied. It is obvious from the figures that the designed control effort has ensured finite time stability of the closed-loop dynamics.

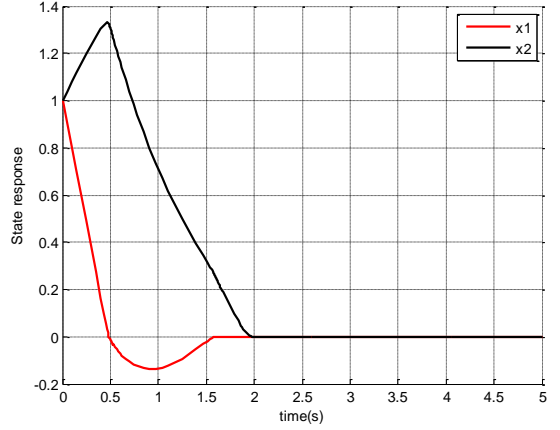

Fig. 3. Closed-loop trajectories example 2

## 8. Application: super-twisting algorithm

In this section, we systematically find Lyapunov function for super-twisting algorithm, which is one of the important algorithms in higher-order sliding mode control. This algorithm is used to attenuate chattering effect and to stabilize sliding surface in finite time sense. In the following, a Lyapunov function ensuring finite time stability is obtained, and then the parameters of this algorithm are designed by using theorem 5 such that the trajectories reach to the origin in finite time. This means that the states are restricted to sliding surface in the sense of finite time.

In this algorithm, the discontinuous part of sliding mode control is constructed as follows:

$$u_1(t) = -2|s|^{\frac{1}{2}}sign(s) - 5s + \int_0^t \left(-10|s|^{\frac{1}{2}}sign(s) - 5s\right)d\tau$$

This part of control signal is designed to guarantee that sliding mode can be maintained (i.e. $s = 0$). Thus, the controlled system obtained as:

$$\dot{s} = -2|s|^{\frac{1}{2}}sign(s) - 5s - 2\int_0^t \left(10|s|^{\frac{1}{2}}sign(s) + 5s\right)d\tau$$

Define the original and slack variables as follows:

$$x_1 = s$$

$$x_2 = y = -\int_0^t \left(10|s|^{\frac{1}{2}}sign(s) + 5s\right)d\tau$$

$$x_3 = |s|^{\frac{1}{2}}sign(s)$$

The problem can be reformulated as to find a Lyapunov function ensuring finite time stability of the system below.

$$\begin{cases} \dot{x}_1 = -2x_3 - 5x_1 + 2x_2 \\ \dot{x}_2 = -10x_3 - 5x_1 \\ \dot{x}_3 = \frac{0.5}{x_1^2}(-2x_3 - 5x_1 + 2x_2) \end{cases}$$

Applying theorem 1 with a given $c = 0.1$ gives the degree 6 Lyapunov function.

Fig.4 shows finite time stability of this algorithm.

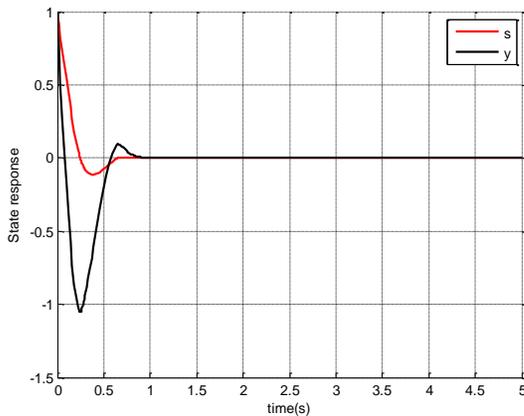

Fig. 4.state response super-twisting algorithm

### 9. Conclusion

A new method for finite time analysis, finite time stabilization, and largest Region of Reaching estimation based on Sum of Squares has been developed in this paper. Several examples were presented to verify applicability of the proposed method. one the important second order sliding mode control (SOSMC), ''super twisting'', also has been designed using presented method in this paper. Benefits of this approach can be summarized as 1) to provide a systematic approach for finite time analysis and also for designing a finite time controller, and 2) to introduce a systematic method for estimating the largest region of reaching 3) existence of efficient numerical methods for solving the problems. For further improvement one can be extend for systems with uncertainties and disturbances. In addition theorems can be extended for discontinuous right-hand sides systems which are studied by differential inclusions. This method also can be extended in order to apply on terminal sliding mode control and consequently apply on the vast applications of this kind of finite time SMC.